\documentclass[aps,prl,twocolumn,superscriptaddress]{revtex4}
\usepackage{amsmath,graphicx,textcomp}
\usepackage{placeins}
\usepackage{booktabs,tabularx}
\usepackage{xcolor}
\usepackage{array}
\newcolumntype{Y}{>{\centering\arraybackslash}X}

\begin{document}            
\title{Interaction between counter-propagating quantum Hall edge channels in the 3D topological insulator BiSbTeSe$_2$ }
\author{Chuan Li}
\thanks{These two authors contributed equally}
\affiliation{MESA+ Institute for Nanotechnology, University of Twente, The Netherlands}
\author{Bob de Ronde}
\thanks{These two authors contributed equally}
\affiliation{MESA+ Institute for Nanotechnology, University of Twente, The Netherlands}
\author{Artem Nikitin}
\affiliation{Van der Waals - Zeeman Institute, Institute of Physics, University of Amsterdam, The Netherlands}
\author{Yingkai Huang}
\affiliation{Van der Waals - Zeeman Institute, Institute of Physics, University of Amsterdam, The Netherlands}
\author{Mark S. Golden}
\affiliation{Van der Waals - Zeeman Institute, Institute of Physics, University of Amsterdam, The Netherlands}
\author{Anne de Visser}
\affiliation{Van der Waals - Zeeman Institute, Institute of Physics, University of Amsterdam, The Netherlands}
\author{Alexander Brinkman}
\affiliation{MESA+ Institute for Nanotechnology, University of Twente, The Netherlands}
\today
\begin{abstract}
The quantum Hall effect is studied in the topological insulator BiSbTeSe$_2$. By employing top- and back-gate electric fields at high magnetic field, the Landau levels of the Dirac cones in the top and bottom topological surface states can be tuned independently. When one surface is tuned to the electron-doped side of the Dirac cone and the other surface to the hole-doped side, the quantum Hall edge channels are counter-propagating. The opposite edge mode direction, combined with the opposite helicities of top and bottom surfaces, allows for scattering between these counter-propagating edge modes. The total Hall conductance is integer valued only when the scattering is strong. For weaker interaction, a non-integer quantum Hall effect is expected and measured.

\end{abstract}

\maketitle
 
The quantum Hall effect (QHE) can be described by the formation of quantized edge state conduction. The conductance of quantum Hall edge modes in a semiconductor is given by $2 n G_0$, where $n$ is the number of modes (linked to the Landau level filling number of the bulk), the multiplication by two is to account for two spins, and $G_0=\frac{e^2}{h}$ is the conductance quantum \cite{Klitzing}. When the electronic dispersion of a material is given by the Dirac equation, the first bulk Landau level sits at the Dirac point and simply provides a conductance contribution of only $\frac{G_0}{2}$, as can be explained by the extra Berry phase of $\pi$ that is obtained in a Landau orbit. For graphene, one then obtains an edge conduction of $4\left(n+\frac{1}{2}\right)G_0$, where the factor of four comes from the twofold spin degeneracy and the twofold orbital degeneracy due to there being Dirac points at the crystallographic K and K' points \cite{Geim}. 

After the discovery of topological insulators, it was soon understood that the Dirac cone of the topological surface state (TSS) of a three-dimensional topological insulator (TI) is not spin degenerate, except at particular Kramer's momenta. Like for graphene, the Berry phase argument provides an offset of $\frac{1}{2}$, and the direction of the conduction channels is determined by the position of the Fermi level in the Dirac cone with respect to the Dirac point (electrons versus holes). Every surface (e.g. top and bottom) then provides an edge conduction of $\left(n+\frac{1}{2}\right)G_0$, rendering the TSS effectively equivalent to one quarter of graphene \cite{Hasanreview}. The top (t) and bottom (b) surfaces of a 3D topological insulator posses Dirac cones of opposite helicities. When the two surfaces are gate-tuned so that the Fermi energy in both systems is either above or below the Dirac point (i.e. two electron or two hole Fermi surfaces), the edge modes of the two surfaces propagate in the same direction, but with opposite helicity. Due to their orthogonality no scattering from one to the other is quantum mechanically allowed. In such a case, the parallel mode conductances add up, yielding an integer quantum Hall effect, i.e. the Hall conductance $G_{xy}=\left(n_t+n_b+1\right)G_0$. This integer quantization has indeed been observed for 3D topological insulators such as BiSbTeSe$_2$ \cite{XuQHE,Purdue}, (Bi$_{1-x}$Sb$_x$)$_2$Te$_3$ \cite{Yoshimi}, HgTe \cite{HgTe}, and magnetically doped topological insulators, where the role of the external magnetic field is replaced by an interal magnetization \cite{Cuizu1,Cuizu2}. 

However, when the top and bottom surfaces of a 3D topological insulator are gate-tuned to different sides of the Dirac point (i.e. one electron and one hole Fermi surface) the edge modes of the two surfaces are counter-propagating, as shown in Fig. 1(a). In this case, the helicities of the states are equal as the sign reversal going from top to bottom surface is cancelled by the sign reversal going from the electron to the hole side of the Dirac cone. This situation is different from the counter-propagating modes in a quantum spin Hall insulator (QSH) \cite{QSH}, where the mode conductance lacks the factor of $\frac{1}{2}$ and where counter-propagating modes at an edge have opposite spins and thus cannot scatter elastically into each other. See the Supplementary Material \cite{supl} for a comparison of the quantum Hall effect in different cases.

\begin{figure*}
	\includegraphics[clip=true,width=16cm]{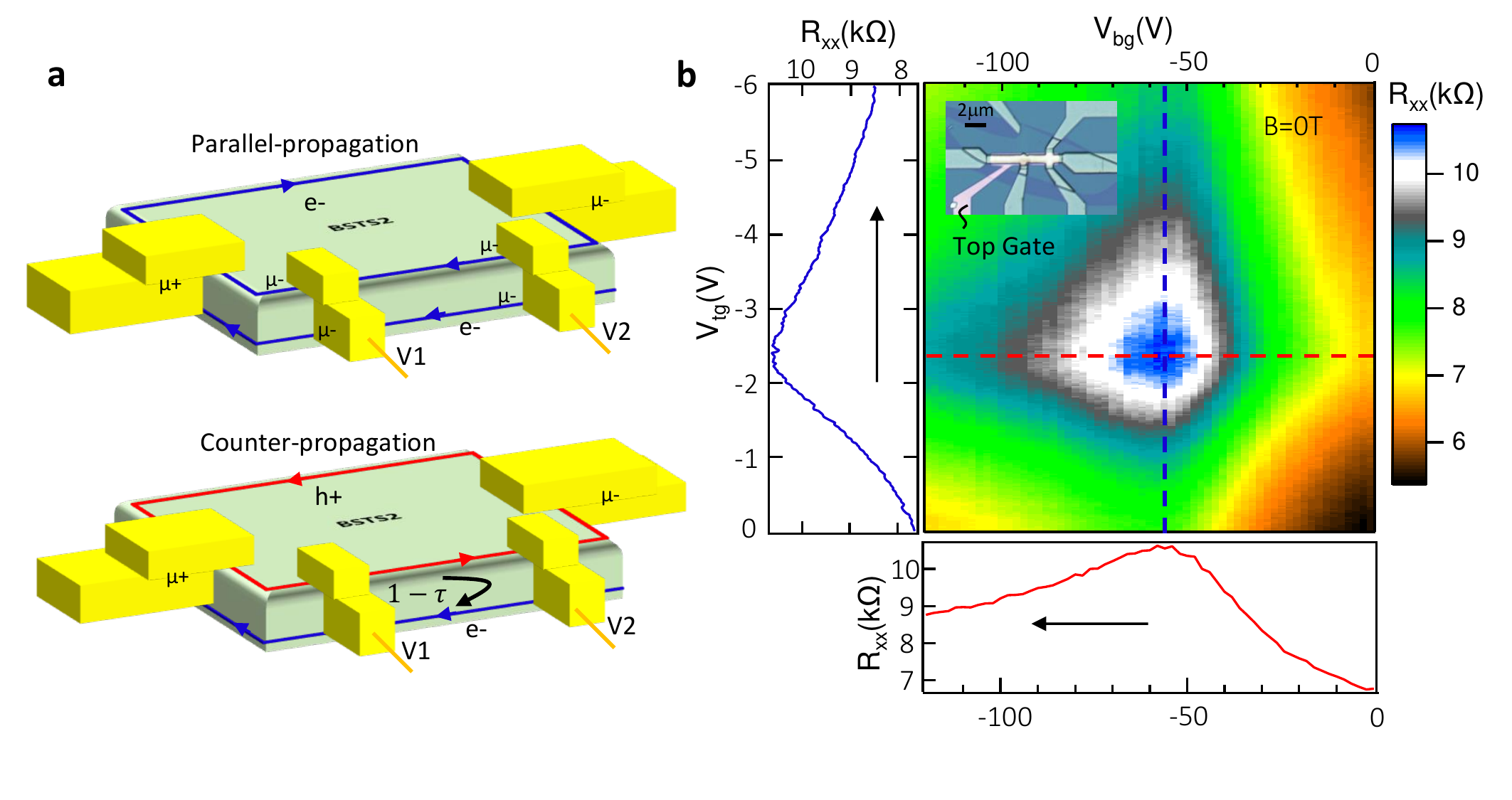}
	\caption{(a) Schematic drawing of a dual-gated quantum Hall device with either parallel propagation or counter-propagation in the edge states of the topological bottom and top surfaces of a 3D topological insulator. In the case of parallel propagation (upper panel), the charge carriers move in the same direction, and the edges of the surfaces form equipotential lines ($\mu_{\pm}$) \cite{Buttiker}. For counter-propagation (lower panel), the electrons and holes come from different potential reservoirs (electrodes) and move in opposite directions. In this case, a non-zero probability exists for backscattering between the top and bottom surfaces, given by ($1-\tau$). (b) Two-dimensional gate map of the longitudinal resistance, $R_{xx}$, as a function of the top and back gate voltages at zero magnetic field. The maximum resistance indicates the chemical potential lying at the the Dirac point. The colour-coded linecuts showing $R_{xx}$ versus gate voltages are also shown, as is an inset showing an optical microscopy image of the device, in which the top gate and the BiSbTeSe$_2$ flake are clearly visible. The black arrows indicate the sweep direction of the measurements.}
	\label{Fig1}
\end{figure*}

Here, we study the interaction between counter-propagating surface states in a three-dimensional topological insulator exploiting independent gate tuning of the upper and lower topological surface states of a BiSbTeSe$_2$ device. We discover non-integer quantum Hall conductance values when the scattering between the surface state modes is suppressed by the use of a large separation between top and bottom surfaces. The non-integer (but rational) conductance values can be understood from the voltage probes being in perfect equilibrium with both the top and bottom edge modes. Modeling the conductance data enables extraction of a value for the probability of scattering between the top and bottom surface modes.

As a three-dimensional topological insulator, stoichiometric BiSbTeSe$_2$ \cite{XuQHE,Neupane,ArakaneBSTS1} is used because of its decent mobility and highly insulating bulk. For growth and details of the e-beam structuring, as well as realization of metal contacts and top and bottom gates, see the Supplementary Material \cite{supl}. Two devices have been characterized at low magnetic fields and both show similar behavior. One device was selected for the high-magnetic field measurements. Fig. 1 depicts the schematic layout of the experiment as well as an optical microscopy image of the device.

We measured the differential resistance for $R_{xx}$ and $R_{xy}$ at zero magnetic field in a Hall-bar-shaped sample while sweeping the top-gate and bottom-gate voltages independently. This gave the data presented in Fig. \ref{Fig1}(b). A single maximum appears in the gate scan range of the map, at which the Fermi levels of both top and bottom surface states are tuned close to their respective Dirac points (DPs). Both top and bottom surfaces were found to be electron doped initially, meaning the DPs of both surfaces are positioned at negative gate voltage. To the left and below the 2D figure, the profiles of $R_{xx}$ as a function of the top (back) gate voltage, $V_{tg}$ ($V_{bg}$), are given for cuts indicated with a blue (red) line. The maximum of the profile as a function of the top-gate voltage does not depend on the back-gate voltage and vice versa, which means that the top and bottom gates only tune their proximate surface states. This shows that the two surfaces are decoupled, split as they are by the insulating bulk, consistent with both previous observations of decoupled BSTS surfaces \cite{Valla} and with the significant thickness chosen for the flake (240 nm). 

\begin{figure*}
	\includegraphics[clip=true,width=15cm]{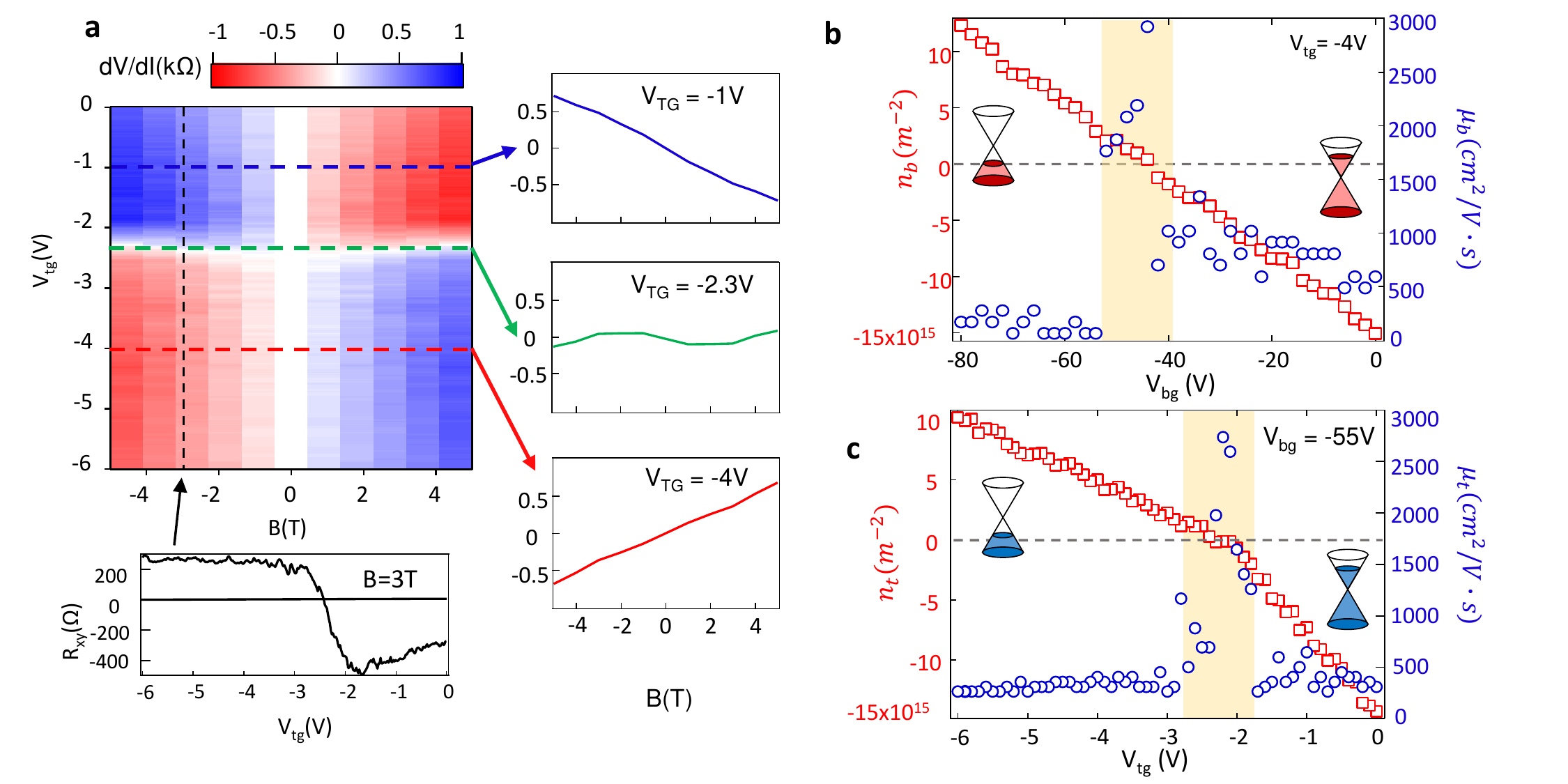}
		\caption{(a) Anti-symmetrized $R_{xy}$ Hall data as a function of top-gate voltage and magnetic field for a fixed back-gate voltage (V$_{bg} =-40 $V). Hall resistance traces as a function of field for different top-gate voltages are shown to the right of the 2D map. The sign of the slope of $R_{xy}$ versus $B$ changes when the top surface is tuned from electron doped (blue) to hole doped (red). At the threshold top-gate voltage (green), the $R_{xy}$ signal is almost flat due to a cancellation of the $R_{xy}^t$ and $R_{xy}^b$ signals. In the bottom figure, a plot of $R_{xy}$ versus $V_{tg}$ is plotted for a field of 2.3 T, and shows a sharp change when $R_{xy}$ crosses zero. (b) Carrier density, $\mathcal{N}_b$, and mobility, $\mu_b$, of the bottom surface as a function of back-gate voltage as the result of a multiband fit to the Hall data. The charge carrier density changes sign as the Fermi level is tuned through the Dirac point. Around the Dirac point (yellow shading), charge puddles give rise to a third conductance contribution. (c) Carrier density, $\mathcal{N}_t$, and mobility, $\mu_t$, of the top surface as a function of top-gate voltage.}
	\label{Fig2}
\end{figure*}

The independent gate tuning capability of the Dirac cones of the two topological surface states is also manifested in the Hall effect data at low magnetic field. Fig. \ref{Fig2}(a) shows the anti-symmetrized $R_{xy}$ Hall signal of top-gate sweeps recorded for different magnetic fields. With the bottom surface slightly electron doped ($V_{bg}= -40$ V), when the top gate crosses the DP, the slope of $R_{xy}(B)$ changes sign, which indicates that we tune the top surface from being electron doped to hole doped. The figure at the bottom shows a sharp change from positive to negative value of $ R_{xy}$ as function of the top-gate voltage.

We deduced the carrier density from the $R_{xy}$ data using a two-band model, in order to account for the top and bottom surface conduction contributions (from the independent gating of the two surfaces, the bulk contribution can be assumed to be negligible). In general, there are four fitting parameters (the top and bottom surface mobilities, $\mu _t$ and $\mu _b$, and the two carrier densities, $\mathcal{N}_t$ and $\mathcal{N}_b$). However, in our case, we benefit from the results of high field measurements (shown and discussed in the Supplementary Material \cite{supl}) to estimate the gate dependence of the carrier density of both surfaces more accurately. This allows us to fix the carrier densities of two surfaces and use the two mobilities as the only fitting parameters. The results of the fitting, see Fig. \ref{Fig2}(b), show that we can tune both the top and bottom surfaces to have very low carrier densities, and thus continuously tune the Fermi level through the DP. When the Fermi level is very close to the DP, rather than needing only two conductance channels, the fitting requires a third contribution. Most likely, this is not due to the side surfaces. In general, the etching steps in the Hall bar fabrication procedure result in a very poor mobility for the side surfaces. Moreover, the side surfaces are oriented parallel to the applied field, meaning that they do not contribute to the $R_{xy}$ signal either. Most likely, the extra contribution arises from spatial charge fluctuations \cite{Martin} in the 2D surface states, also observed for BiSbTeSe$_2$ \cite{Borgwardt}. From the multi-band fit, the carrier density of these charge puddles is estimated to be about $5\times 10^{15}$ m$^{-2}$. 

\begin{figure*}
	\includegraphics[clip=true,width=10cm]{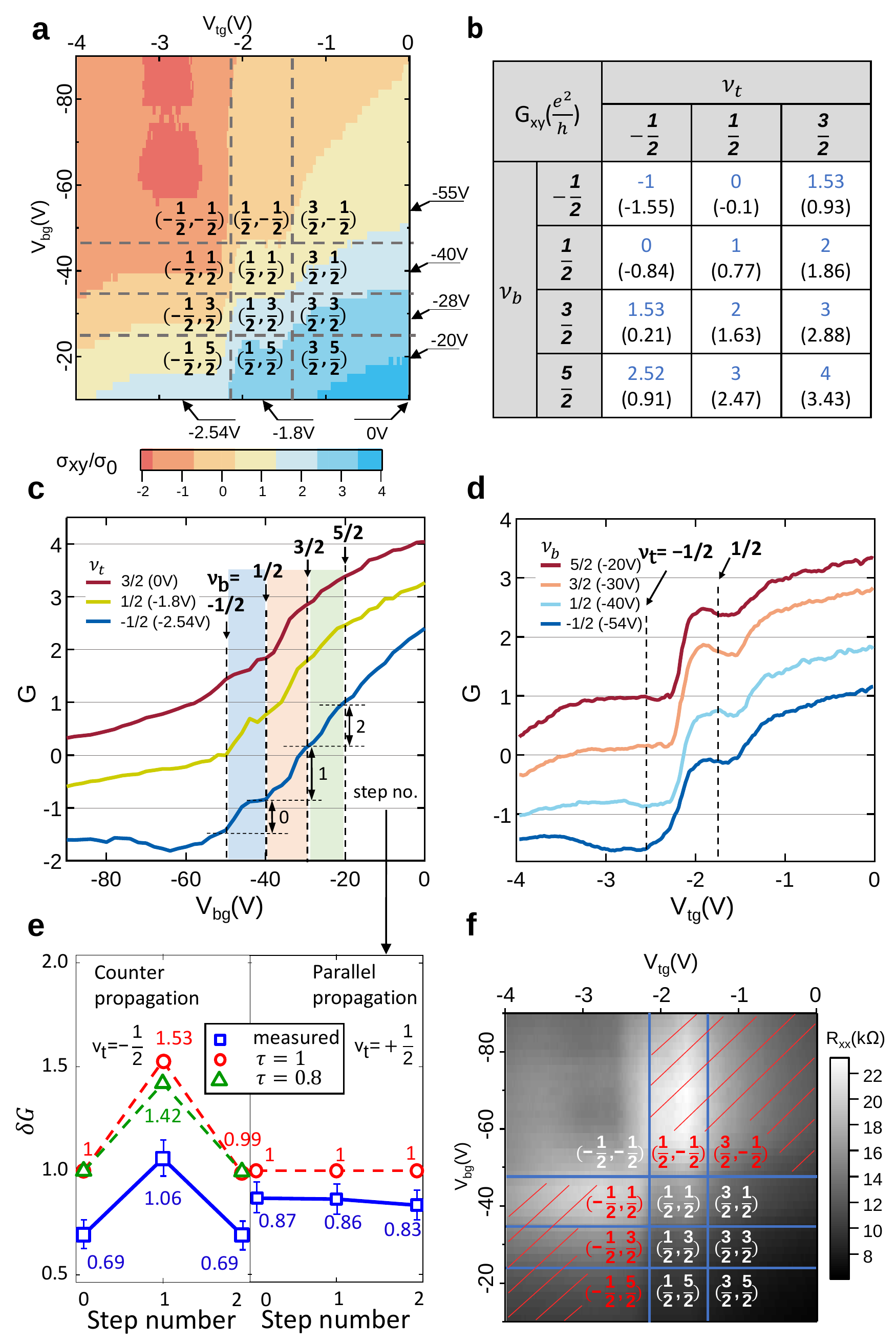}
	\caption{(a) Hall conductance ($G_{xy}$) as a function of top-gate voltage ($V_{tg}$) and back-gate voltage ($V_{bg}$) at 15 T. The dashed lines indicate the borders between plateaus corresponding to different filling factors, noted as $(\nu_t,\nu_b)$. The voltage indicators on the outside of the frame mark the positions of the cross-sections in the subsequent panels. (b) Expected values of the Hall conductance for different filling factors in units of $\frac{e^2}{h}$. The measured values are shown between brackets. (c) Back-gate voltage dependence of $G_{xy}$ at the three values of $\nu_t$ shown in (b). The vertical, dashed lines indicate different $\nu_b$. (d) Top-gate voltage dependence of $G_{xy}$ at the four different $\nu_b$'s given in (b). The vertical, dashed lines now indicate different $\nu_t$. (e) Measured and calculated values for the change in $G_{xy}$ for successive bottom surface Landau levels, as indicated by the steps in (c) when going from one $\nu_b$ value to the next. Error bars are extracted from the averaging carried out of the gating map data. Left: $\nu_t=-\frac{1}{2}$ (counter-propagating modes). Right: $\nu_t=\frac{1}{2}$ (parallel propagation). The calculated step sizes are shown both for perfect transmission ($\tau =1$) and for $\tau=0.8$. (f) Longitudinal resistance, $R_{xx}$, at $B=15$ T versus $V_{tg}$ and $V_{bg}$. Blue solid lines indicate the borders of the Landau level plateaus. The hatched areas indicates regions with counter-propagating modes.}
	\label{Fig3}
\end{figure*}

At high magnetic field, the two gates can be used to tune the Fermi level between different Landau levels (LLs). A gate map of $G_{xy}$ at 15 T and 50 mK is shown in Fig. \ref{Fig3}(a). We deduced $G_{xy}$ from the measured $R_{xx}$ and $R_{xy}$ by inversion of the resistivity tensor, after we symmetrized the $R_{xx}$ data and anti-symmetrized the $R_{xy}$ data to minimize possible geometric effects. Despite the fact that the quantization of the LLs is not perfect at only 15 T, we can already see that the $G_{xy}$ gate map is divided into several quasi-rectangular areas. These plateaus correspond to different filling factor combinations of $\nu_t=n_t+\frac{1}{2}$ and $\nu_b=n_b+\frac{1}{2}$, which we have indicated in brackets in the figure. 

To get a better understanding of the expected quantized Hall conductance in this combined system of two surface states, we modeled the system using the Landauer-B\"{u}ttiker formula (see Supplementary Material \cite{supl} for the modeling details). We theoretically expect an unusual non-integer Hall conductance in the regime for which the two surfaces are populated by charge carriers of opposite sign (lower panel of Fig. \ref{Fig1}(a)), but equal helicity. Intuitively, when the coupling between electrons and holes is strong, the counter-propagating states counteract each other and will cancel when summing the Hall conductances, but this picture only holds when the counter-propagating filling factors are exactly opposite (e.g. $\nu_t=-\nu_b=\frac{1}{2}$). In general, counter-propagating edge modes start off from different current injection electrodes, and have, therefore, different chemical potentials, see also Fig. 1(a). If there is no interaction possible between the surface channels through the bulk, the only way to get equilibrium is to equalize the potential inside the metal electrodes. Using this as a boundary condition, we theoretically expect non-integer values for the Hall conductances in the counter-propagating regime, even for perfect transmission of the edge channels. The calculated and measured values are shown in Fig. \ref{Fig3}(b). Cross-sectional cuts of the data are shown in Figs. \ref{Fig3}(c) and \ref{Fig3}(d).

Due to the imperfect edge channels at these moderate magnetic fields (i.e. $\mu B \ll 1$ not being fulfilled), the values for the Hall conductance deviate from the expected values, and $R_{xx}$ does not go completely to zero. This effect becomes most apparent for the top-gate dependence of $G_{xy}$ at constant bottom surface Landau level, as shown in Fig. \ref{Fig3}(d), and is strongest in the regimes for which the surfaces are populated by charge carriers of opposite sign (we note that the mobility of the holes is generally lower than the mobility of the electrons in topological insulators, consistent with our observation in Fig. 2). Indeed, impurities or defects on the side surfaces of a topological insulator are predicted to lead to hybridization of the edge states of the two surfaces \cite{ZhangsideSS}. 

However, the bottom surface shows better quantization values (perhaps due to better protection during device processing), hence we use Fig. 3(c) rather than Fig. 3(d) for the subsequent analysis. Especially when we look at the change in the Hall conductance at constant $\nu_t$ when going from one $\nu_b$ to the other, a quantitative analysis can be made as regards the nature of the coupling between counter-propagating edge modes. We plot both the calculated (red, green) and the measured (blue) Hall conductance changes ($\delta G_{xy}$) when changing $\nu_b$ in Fig. \ref{Fig3}(e). The step numbers 0, 1 and 2 correspond, respectively, to $\nu _b=-\frac{1}{2} \rightarrow \frac{1}{2}$, $\frac{1}{2}\rightarrow\frac{3}{2}$, and $\frac{3}{2}\rightarrow\frac{5}{2}$. Note, that all experimental values are lower than theoretically expected, likely due to the non-vanishing shunting conductance of the bulk states. Despite the overall lowering factor, for counter-propagating modes a clearly non-monotonic change in the Hall conductance is experimentally observed around step 1 for $\nu _t =-\frac{1}{2}$, as predicted by the model. On the other hand, the Hall conductance change stays almost constant for parallel propagation, when $\nu_t=\frac{1}{2}$, also in line with the model. 

This observation is different from previous reports on topological insulators \cite{XuQHE,Purdue,HgTe}, where the total Hall conductance remained integer valued, even in the case of counter-propagating modes. We note that our devices have a significantly larger separation between the surfaces and that the scattering between counter-propagating modes is therefore reduced. We model the coupling between counter-propagating modes with an effective mode transmission probability, $\tau$. Then, the probability of reflecting into the mirrored, counter-propagating channel (both opposite charge \textit{and} propagation direction) is $1-\tau$. When $\tau = 1$, the counter-propagating channels are only coupled through the equilibration of the chemical potential  of the edge modes inside the voltage probe electrodes. However, when $\tau =0$, the counter-propagating channels are fully coupled, and the Hall conductance is found to be integer valued (see Supplementary Material \cite{supl} for details). This is most likely the explanation of the integer quantum Hall effect seen in thinner samples. The non-monotonic change in Hall conductance observed in our case is consistent with a large value of $\tau$ (for comparison, also the expected values for $\tau=0.8$ are shown in Fig. 3(e), which resemble the experimentally observed relative step heights well), as expected for thicker flakes.
 
Interestingly, the longitudinal resistance, $R_{xx}$, also behaves differently for parallel propagation and counter-propagating edge modes. For parallel propagation (areas without hatching in Fig. \ref{Fig3}(f)), $R_{xx}$ would tend to zero if the edge modes were to become increasingly ideal at higher magnetic field. However, if the two topological surfaces have counter-propagating edge states (hatched regions in Fig. \ref{Fig3}(f)), $R_{xx}$ becomes large. We calculated $R_{xx}$ using the Landauer-B\"{u}ttiker formula (see Supplementary Material \cite{supl}). For $\nu_t=\pm1/2$ and $\nu_b = \mp1/2$, we find $\rho_{xx}=\frac{h}{\tau e^2}$. If the channels are very transparent ($\tau\approx1$), $R_{xx}$ should be approximately $G_0^{-1}$, which can be understood from the equilibration of the chemical potential in the voltage probe electrodes. This situation to also applicable to observations in the HgTe/CdTe quantum spin Hall state, where $\tau=1$ because of the opposite spin of the modes \cite{QSH}, albeit with a factor of two difference because of the different Berry phase.  If $\tau\ll 1$, the two counter-propagating channels are strongly coupled, since the backscattering rate is high, so $R_{xx}$ is expected to be large. The gate map of $R_{xx}$ at 15 T is shown in Fig. 3(f). The filling factors for both surfaces are indicated using the notation ($\nu_t, \nu_b$). It can be seen that both $R_{xx}(\frac{1}{2},-\frac{1}{2})=22.5$ k$\Omega$ and $R_{xx}(-\frac{1}{2},\frac{1}{2})=20.5$ k$\Omega$ are close to $G_0^{-1}$, again indicative of $\tau$ being close to 1. For the thinner sample of Xu \textit{et al.} \cite{XuQHE}, based on their measured value for $R_{xx}$, we estimate $\tau= 0.1$, which is indeed an order of magnitude smaller, indicating more proximate and thus more strongly coupled edge channels, fully consistent with their observation of an integer quantum Hall effect.

In conclusion, the Fermi level has been controlled independently for the upper and lower surface states of a 3D topological insulator using a dual-gating configuration. The developing quantum Hall states are observed at a magnetic field of 15 T. Applying the Landauer-B\"{u}ttiker formalism, we simulate the system for both a parallel and counter propagation edge state configuration and we experimentally confirm a non-monotonic change in the Hall conductance for counter-propagating states when compared to the integer quantum Hall effect. Our data show that it is the interaction between counter-propagating modes that results in the non-integer quantum Hall effect. The interaction can be understood from the equilibration of the chemical potential in the electrodes and the scattering between the edge modes of the top and bottom surfaces.

Compared to the well studied electron-hole quantum Hall bilayers in semiconducting 2D heterostructures (e.g see \cite{Mendez1985,e-hsyst1,Eisenstein}), the topological surface states hold up the intriguing prospect of showing fractional exchange statistics, when combined with superconductivity, due to the helical nature of the edge modes. Counter-propagating and spin-resolved edge modes have also been realized in quantum spin Hall insulators \cite{QSH} and twisted bilayer graphene \cite{JavierPRL2012gra,SchanezFracQH}, but scattering between counter-propagating edge modes, as reported here, is only possible for 3D topological insulators, providing an additional control parameter in quantum Hall experiments and applications. The combination of edge mode interaction and potential equilibration in the electrodes might also be a suitable platform to investigate models for scattering in the fractional quantum Hall effect \cite{Kane} and independent tuning of quantum Hall edge states by the magnetic proximity effect \cite{Moodera1,Moodera2,Moodera3}.

This work was financially supported by the Netherlands Organization for Scientifc Research (NWO), and the European Research Council (ERC) through a Consolidator Grant.


\begin{thebibliography}{1}
\bibitem{Klitzing} K. Von Klitzing, G. Dorda, M. Pepper, Phys. Rev. Lett. \textbf{45}, 494 (1980).
\bibitem{Geim} K.S. Novoselov, Z. Jiang, Y. Zhang, S.V. Morozov, H.L. Stormer, U. Zeitler, J.C. Maan, G.S. Boebinger, P. Kim, A.K. Geim, Science \textbf{315}, 1379 (2007).
\bibitem{Hasanreview} M.Z. Hasan, C.L. Kane, Rev. Mod. Phys. \textbf{82}, 3045 (2010).
\bibitem{XuQHE} Y. Xu, I. Miotkowski, C. Liu, J. Tian, H. Nam, N. Alidoust, J. Hu, C.K. Shih, M.Z. Hasan, Y.P. Chen. Nature Phys. \textbf{10}, 956 (2014).
\bibitem{Purdue} Y. Xu, I. Miotkowski, Y.P. Chen, Nature Comm. \textbf{7}, 11434 (2016).
\bibitem{Yoshimi} R. Yoshimi, A. Tsukazaki, Y. Kozuka, J. Falson, K.S. Takahashi, J.G. Checkelsky, N. Nagaosa, M. Kawasaki, Y. Tokura, Nature Comm. \textbf{6}, 6627 (2015).
\bibitem{HgTe} C. Br\"une, C.X. Liu, E.G. Novik, E.M. Hankiewicz, H. Buhmann, Y.L. Chen, X.L. Qi, Z.X. Shen, S.C. Zhang, L.W. Molenkamp, Phys. Rev. Lett. \textbf{106}, 126803 (2011).
\bibitem{Cuizu1} C.Z. Chang \textit{et al.}, Science \textbf{340}, 167 (2013).
\bibitem{Cuizu2} C.Z. Chang, W. Zhao, D.Y. Kim, H. Zhang, B.A. Assaf, D. Heiman, S.C. Zhang, C. Liu, M.H.W. Chan, J.S. Moodera, Nature Mater. \textbf{14}, 473 (2015). 
\bibitem{QSH} M. K\"onig, S. Wiedmann, C. Br\"une, A. Roth, H. Buhmann, L.W. Molenkamp, X.L. Qi, and S.C. Zhang, Science \textbf{318}, 766 (2007).
\bibitem{supl} Supplementary Material.
\bibitem{Buttiker} M. B\"uttiker, Phys. Rev. B \textbf{38}, 9375 (1988).
\bibitem{Neupane} M. Neupane, S.Y. Xu, L.A. Wray, A. Petersen, R. Shankar, N. Alidoust, C. Liu, A. Fedorov, H. Ji, J.M. Allred, Y.S. Hor, T.R. Chang, H.T. Jeng, H. Lin, A. Bansil, R.J. Cava, and M.Z. Hasan, Phys. Rev. B \textbf{85}, 235406 (2012).
\bibitem{ArakaneBSTS1} T. Arakane, T. Sato, S. Souma, K. Kosaka, K. Nakayama, M. Komatsu, T. Takahashi, Z. Ren, K. Segawa, Y. Ando, Nature Comm. \textbf{3}, 636 (2012).
\bibitem{Valla}  V. Fatemi, B. Hunt, H. Steinberg, S.L. Eltinge, F. Mahmood, N.P. Butch, K. Watanabe, T. Taniguchi, N. Gedik, R.C. Ashoori, P. Jarillo-Herrero, Phys. Rev. Lett. \textbf{113}, 206801 (2014).
\bibitem{Martin} J. Martin, N. Akerman, G. Ulbricht, T. Lohmann, J.H. Smet, K. von Klitzing, A. Yacoby, Nature Phys. \textbf{4}, 144 (2007).
\bibitem{Borgwardt} N. Borgwardt \textit{et al.}, Phys. Rev. B \textbf{93}, 245149 (2016). 
\bibitem{ZhangsideSS} Y.Y. Zhang, X.R. Wang, X.C. Xie, J. Phys. Cond. Matter \textbf{24}, 15004 (2012).
\bibitem{Mendez1985} E.E. Mendez, L. Esaki, L.L. Chang, Phys. Rev. Lett. \textbf{55}, 2216 (1985).
\bibitem{e-hsyst1} K. Suzuki, S. Miyashita, Y. Hirayama, Phys. Rev. B \textbf{67}, 195319 (2003).
\bibitem{Eisenstein} J.P. Eisenstein, A.H. MacDonald, Nature \textbf{432}, 691 (2004).
\bibitem{JavierPRL2012gra} J.D. Sanchez-Yamagishi, T. Taychatanapat, K. Watanabe, T. Taniguchi, A. Yacoby, P. Jarillo-Herrero, Phys. Rev. Lett. \textbf{108}, 076601 (2012).
\bibitem{SchanezFracQH} J.D. Sanchez-Yamagishi, J.Y. Luo, A.F. Young, B. Hunt, K. Watanabe, T. Taniguchi, R.C. Ashoori, P. Jarillo-Herrero, arXiv/1602.06815 (2016).
\bibitem{Kane} C.L. Kane, M.P.A. Fisher, Phys. Rev. B \textbf{51}, 13449 (1995).
\bibitem{Moodera1} M. Li \textit{et al.}, Phys. Rev. Lett. \textbf{115}, 087201 (2015).
\bibitem{Moodera2} Z. Jiang, C.Z. Chang, C. Tang, P. Wei, J.S. Moodera, J. Shi, Nano Lett. \textbf{15}, 5835 (2015).
\bibitem{Moodera3} F. Katmis \textit{et al.}, Nature \textbf{533}, 513 (2016).
\end{thebibliography}
\end{document}


\title{Interaction between counter-propagating quantum Hall edge channels in the topological insulator BiSbTeSe$_2$\\ - Supplemental Material}
\author{Chuan Li}
\thanks{These two authors contributed equally}
\affiliation{MESA+ Institute for Nanotechnology, University of Twente, The Netherlands}
\author{Bob de Ronde}
\thanks{These two authors contributed equally}
\affiliation{MESA+ Institute for Nanotechnology, University of Twente, The Netherlands}
\author{Artem Nikitin}
\affiliation{Van der Waals - Zeeman Institute, Institute of Physics, University of Amsterdam, The Netherlands}
\author{Yingkai Huang}
\affiliation{Van der Waals - Zeeman Institute, Institute of Physics, University of Amsterdam, The Netherlands}
\author{Mark S. Golden}
\affiliation{Van der Waals - Zeeman Institute, Institute of Physics, University of Amsterdam, The Netherlands}
\author{Anne de Visser}
\affiliation{Van der Waals - Zeeman Institute, Institute of Physics, University of Amsterdam, The Netherlands}
\author{Alexander Brinkman}
\affiliation{MESA+ Institute for Nanotechnology, University of Twente, The Netherlands}
\today
\begin{abstract}
1. Quantum Hall effect comparison 

2. Sample fabrication

3. Gate-dependent hysteresis

4. Determining the Landau levels

5. Multiband fitting results

6. Landau level spacing and effective dielectric constant

7. Landauer-B\"uttiker formalism for topological insulator quantum Hall edge states   

8. (Anti-)symmetrization of $ \rho_{xx} $ and $\rho_{xy}$ signals in the dissipative quantum Hall regime
\end{abstract}

\maketitle
 
\renewcommand{\thefigure}{S\arabic{figure}}   
\renewcommand{\figurename}{Fig.}

\subsection{1. Quantum Hall effect comparison}
In this section, an overview is given over the different versions of the quantum Hall effect in different materials, and the possible values for the quantized mode conductances in the different cases, see also Fig. S1. A spin-degenerate semiconductor has two  conductance quanta per mode because of spin. The modes run in the direction as dictated by the magnetic field. Graphene has an additional degeneracy factor of two because of the valley degeneracy. The Dirac type of dispersion in graphene has an associated pseudospin-momentum locking that provides an offset of $\frac{1}{2}$. This zeroth Landau level is also refered to as a consequence of the non-zero Berry phase in a Landau orbit. The $\frac{1}{2}$-term is an exciting illustration of how a non-trivial $Z_2$ topological number appears in the quantum Hall language of topological TKKN quantum numbers. The quantum spin Hall insulator lacks this factor of $\frac{1}{2}$. The quantum spin Hall insulator has counterpropagating modes, hence conductance in both directions. The spin is opposite for the two directions, hence the modes are quantum mechanically orthogonal and no elastic scattering between the modes is allowed. For the three-dimensional topological insulator, the mode direction is determined by the magnetic field and the nature of the carriers (electrons when the chemical potential lies above the Dirac point, holes when the chemical potential is below the Dirac point). Co-propagating modes have opposite spin (because the helicity is reversed for the top and bottom surfaces) and a factor of $\frac{1}{2}$ because of the helical spin-momentum locking in the Dirac cone. Counter-propagating modes have the same spin and scattering between the modes is quantum mechanically allowed. The transparency of the mode, $T$, can then be smaller than one.

\begin{figure}
	\includegraphics[clip=true,width=10cm]{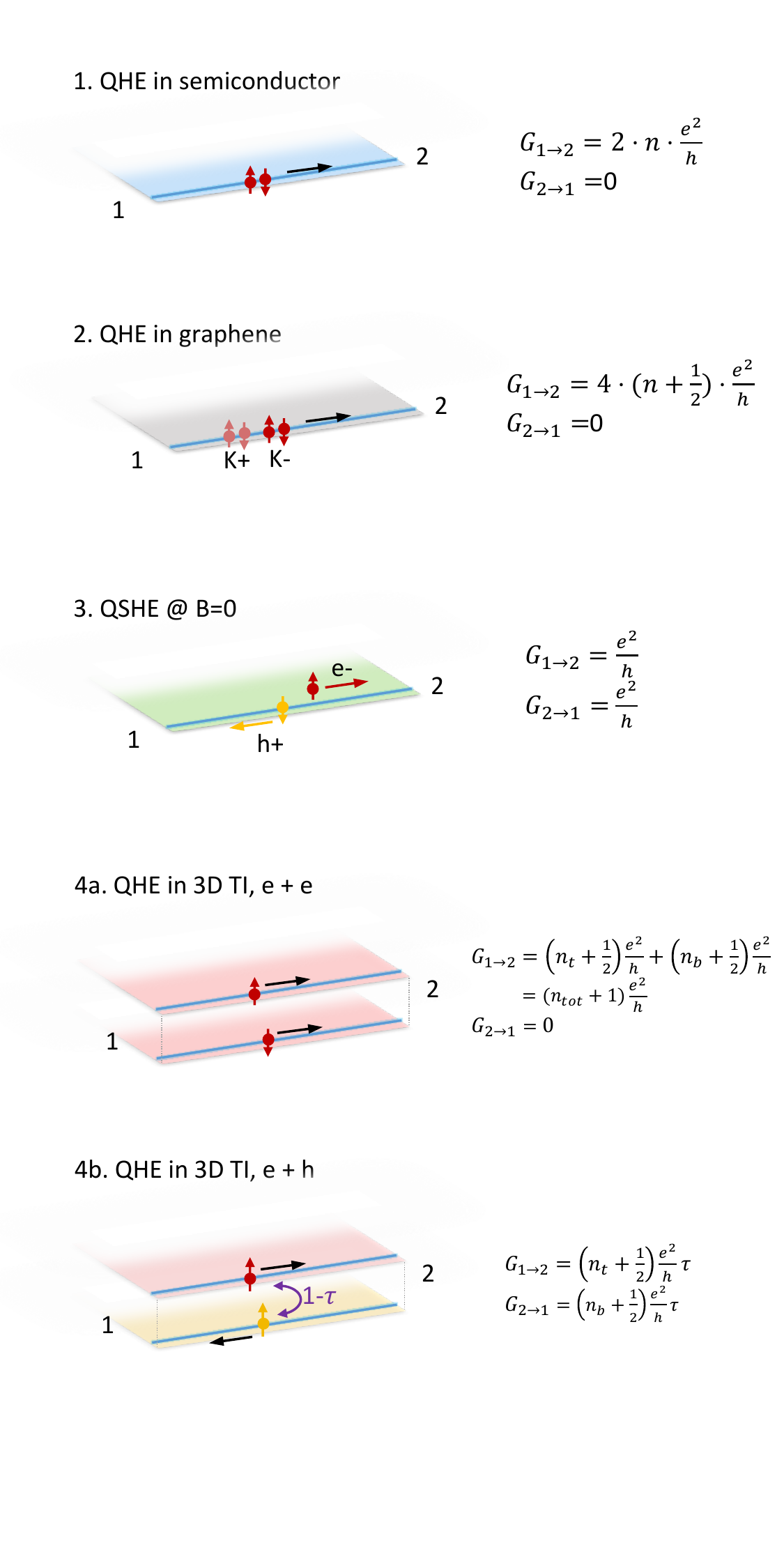}
	\caption{The quantum Hall effect and mode conductances are shown for the following four cases: 1. degenerate semiconductor, 2. graphene, 3. quantum spin Hall insulator (i.e. 2D topological insulator at zero magnetic field), 4. 3D topological insulator with co- and counter-propagating edge modes.}
	\label{Fig:QHE}
\end{figure} 

\subsection{2. Sample fabrication}
High quality BiSbTeSe$_2$ single crystals were grown using a modified Bridgman method. Stoichiometric amounts of the high purity elements Bi (99.999\%), Sb (99.9999\%), Te (99.9999\%) and Se (99.9995\%) were sealed in an evacuated quartz tube and placed vertically in a tube furnace. The material was kept at 850 \textcelsius\ for three days and then cooled down to 500 \textcelsius\ with a speed of 3 \textcelsius\ per hour, followed by cooling to room temperature at a speed of 10 \textcelsius\ per minute. We exfoliated single crystal flakes onto a highly doped silicon substrate topped with a 300 nm thick SiO$_2$ layer on top. Nb/Pd (80/10 nm) metal contacts are fabricated using sputter deposition and e-beam lithography. After making the contacts, we shaped the flakes into a Hall bar structure using e-beam lithography and Ar$^{+}$ etching. Next, the entire central area of the BiSbTeSe$_2$ flake is covered with a 20 nm thick Al$_2$O$_3$ layer using atomic layer deposition at 100 \textcelsius. In the final step, the top gate is realized by using e-beam lithography and lift-off of a sputter deposited Au layer. Two devices have been characterized at low magnetic fields and both show similar behavior. One device was selected for the high-magnetic field measurements. Fig. 1 depicts the schematic layout of the experiment as well as an optical microscopy image of the device.

\subsection{3. Gate-dependent hysteresis}
\begin{figure}
	\includegraphics[clip=true,width=16cm]{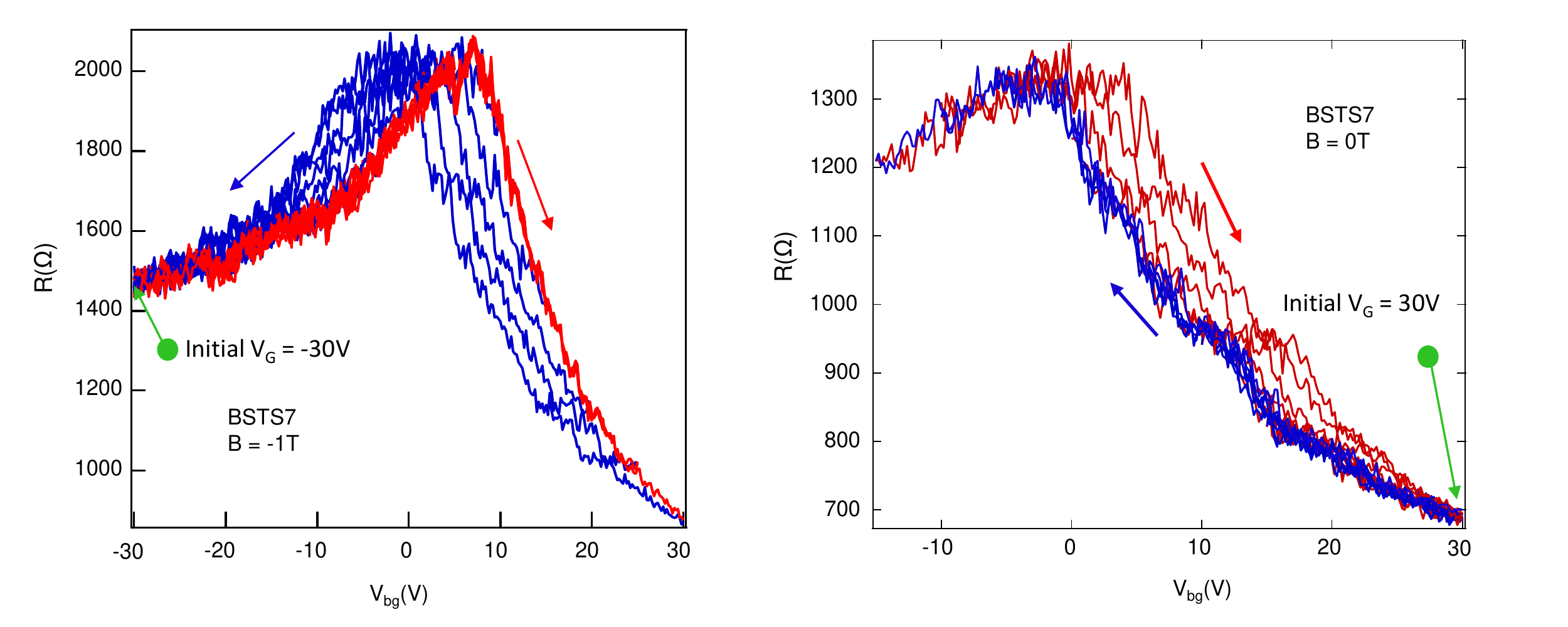}
	\caption{ (a) Gate-dependent hysteresis is shown, where red indicates increased gate voltage and blue decreased gate voltage. Each up-sweep starts from $V_G = -30V$ and ends at different voltages (+30V, 20V, $\cdots$ -20V). The down sweeps always come back to $V_G = -30V$. Since the up sweeps all start from the same voltage (-30V), each of the six profiles measured retrace each other. The down sweeps in such a cycle always start from a different voltage, which therefore produces clear gate-dependent hysteresis loops. (b) All the down sweeps have the same initial starting point ($V_G=30V$) and end at different values before sweeping back up. Now the hysteresis is seen in the up sweeps.}
	\label{Fig:Fig1s-hysteresis}
\end{figure}

For both surfaces, gate-dependent sweeps show hysteretic behavior. The hysteresis is probably mainly due to trapped charges in bulk defects, which do not contribute to the transport. By carefully comparing the data of up and down sweeps, we find that the curves are fully reproducible for identical initial conductions and sweeping direction. In Fig. \ref{Fig:Fig1s-hysteresis}, we show a series of representative gate voltage sweeps. All curves with the same initial state, and the same sweeping direction exactly retrace. This memory-like behavior is observed consistently in all of our samples for both top and bottom gating. 

\subsection{4. Determining the Landau levels}
This section deals with how one can determine the filling factors for each surface if the LLs are not fully quantized. First, we use our low field results. We determine the gate voltage, $V_G$, at which the charge carrier density is the lowest by changing only one gate voltage. Normally, this point corresponds to the maximum of $R_{xx}(V_G)$. Once the Fermi level crosses the Dirac point, the $R_{xy}$ signal changes sign and a large peak or dip is observed in the measured differential resistance. Such a significant change in $R_{xy}$ signal gives a clear indication of the $\nu=\frac{1}{2}$ and $\nu=-\frac{1}{2}$ LLs (depending on the sign of the gate voltage). In this way, we located four levels at $V_{bg}=$-40V and -55V for the $\nu_b = \pm\frac{1}{2}$, and $V_{tg}=-2$V and -3V for $\nu_t = \pm\frac{1}{2}$ respectively. Here, the labels t [b] represent the top [bottom] surface. After assigning these filling factors, we can determine the filling factors of the other squares in the $G_{xy}$ gate map, shown in the main paper in Fig. 3(a).

\FloatBarrier
\subsection{5. Multiband fitting results}

\begin{figure}
	\includegraphics[clip=true,width=16cm]{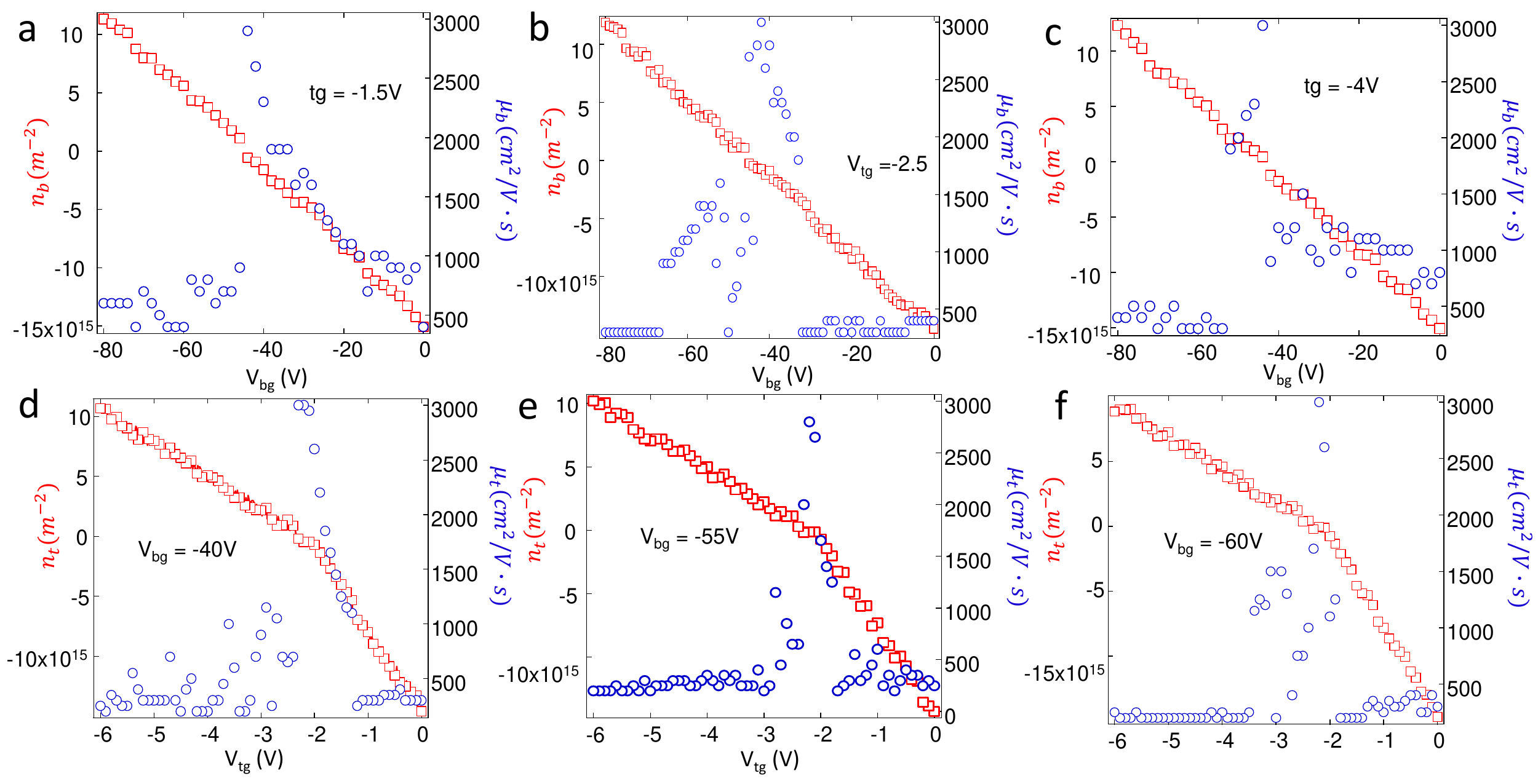}
	\caption{(a)-(c) Bottom-gate dependence of the carrier density, $n_b$, and mobility, $\mu_b$, for different top-gate voltages, as resulting from a multiband fit to the Hall data. (d)-(f) Top-gate dependence of the carrier density, $n_t$, and mobility, $\mu_t$, for different bottom-gate voltages. The mobilities increase near the Dirac point.}
	\label{Fig:Fig2s-fitting}
\end{figure}

For the fitting of the low-field Hall data, as presented in Fig. 2 of the main text, a two- or three-band model was used. The $R_{xy}$ curves and the zero field longitudinal resistance values were used to obtain the carrier density and mobility values for both the top and bottom surfaces. The third band that was used to fit the data is ascribed to charge puddles and only plays a role when $E_F$ is close to $E_{Dirac}$. In Fig. \ref{Fig:Fig2s-fitting}, we show the fitting results for different gate voltages.

As a guideline for the fitting, the carrier density of the gated surface was only allowed to change linearly with the gate voltage. The rate of change was determined using the effective dielectric constants obtained from the LL fan diagrams discussed in Supplementary Section 5 below. For the other surface and the charge puddles, a small range in carrier density was chosen such that the data could be well fitted across the gate voltage range. However, the mobility of the gated surface was allowed to vary distinctly more, since the mobility increases significantly close to the Dirac point, as shown in Fig. S3.

It is evident that the bulk conductivity does not appear to be relevant in the data fitting. This observation was already warranted from the independent gate tuning of the top and bottom surfaces. We argue here that bulk conductivity is also negligible when it comes to equilibration between electrodes. The independent gating, as shown in main text Fig. 1(b), implies a negligible conductance between the top and bottom surfaces on the scale of the longitudinal conductance. Given the order of magnitude of the longitudinal resistance, $R_{xx}$ of 10 k$\Omega$, and the dimensions of the Hall bar (i.e. a flake thickness of 240 nm, a width of about 1 $\mu$m and a length of more than 5 $\mu$m), the bulk resistivity is then found to be much larger than 2 $\Omega$cm. The bulk channel resistance between electrodes (spacing is 2 $\mu$m) is then found to be much larger than 2$\times 10^5$ $\Omega$, meaning that it can be neglected when compared to the quantum of the resistance.

\begin{figure}
	\includegraphics[clip=true,width=17cm]{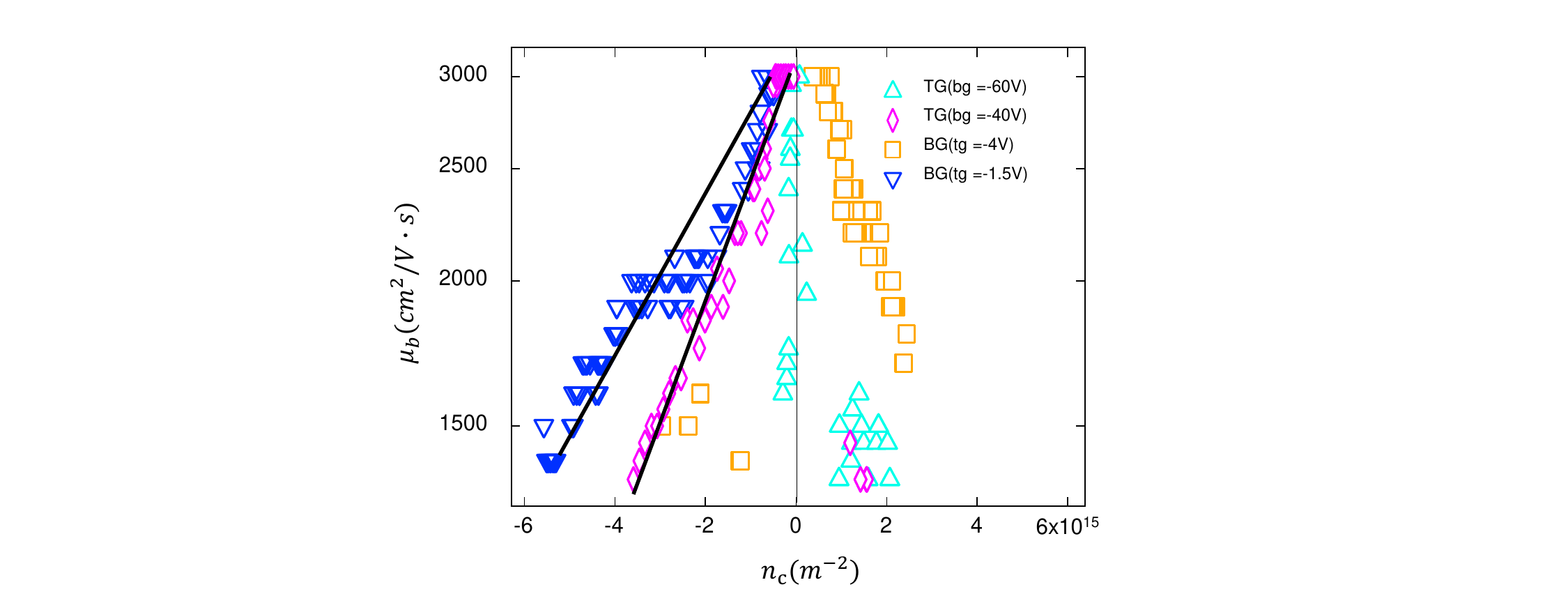}
	\caption{The mobility exponentially decays with increase of the carrier density. We show exemplary fits of one top- and one bottom-gated datasets using the relation $\mu =\mu_0 e^{-n_c/n_0}$. The fitted $n_0$ corresponds to the carrier density of the fixed surface (with constant gate voltage).}
	\label{Fig:Fig3s-mobvsNc}
\end{figure} 

The increased mobility close to the Dirac point is similar to what has been reported for (Bi$_{0.04}$Sb$_{0.96}$)$_2$Te$_3$ \cite{TianSciRep2014}, BiSbTeSe$_2$ (see supplementary material of Ref. \onlinecite{XuQHE}), and also for graphene \cite{HighMobsuspendedGraphene}. Multiple scattering mechanisms can be responsible for such a carrier density dependent mobility \cite{GauravSciRep2014,DohunPRL2012}: defects on the topological insulator surfaces, phonons, or even edge roughness. Experimentally, near the Dirac point, we observe a logarithmic dependence of the mobility on the carrier density with different pre-factors that depend on the gate voltage used. The observed kink in the dependence of the top-surface carrier density on top-gate voltage must relate to a change in gating efficiency across the Dirac point, likely related to a change in dielectic screening.

\FloatBarrier
\subsection{6. Landau level spacing and effective dielectric constant}
\begin{figure}
	\includegraphics[clip=true,width=17cm]{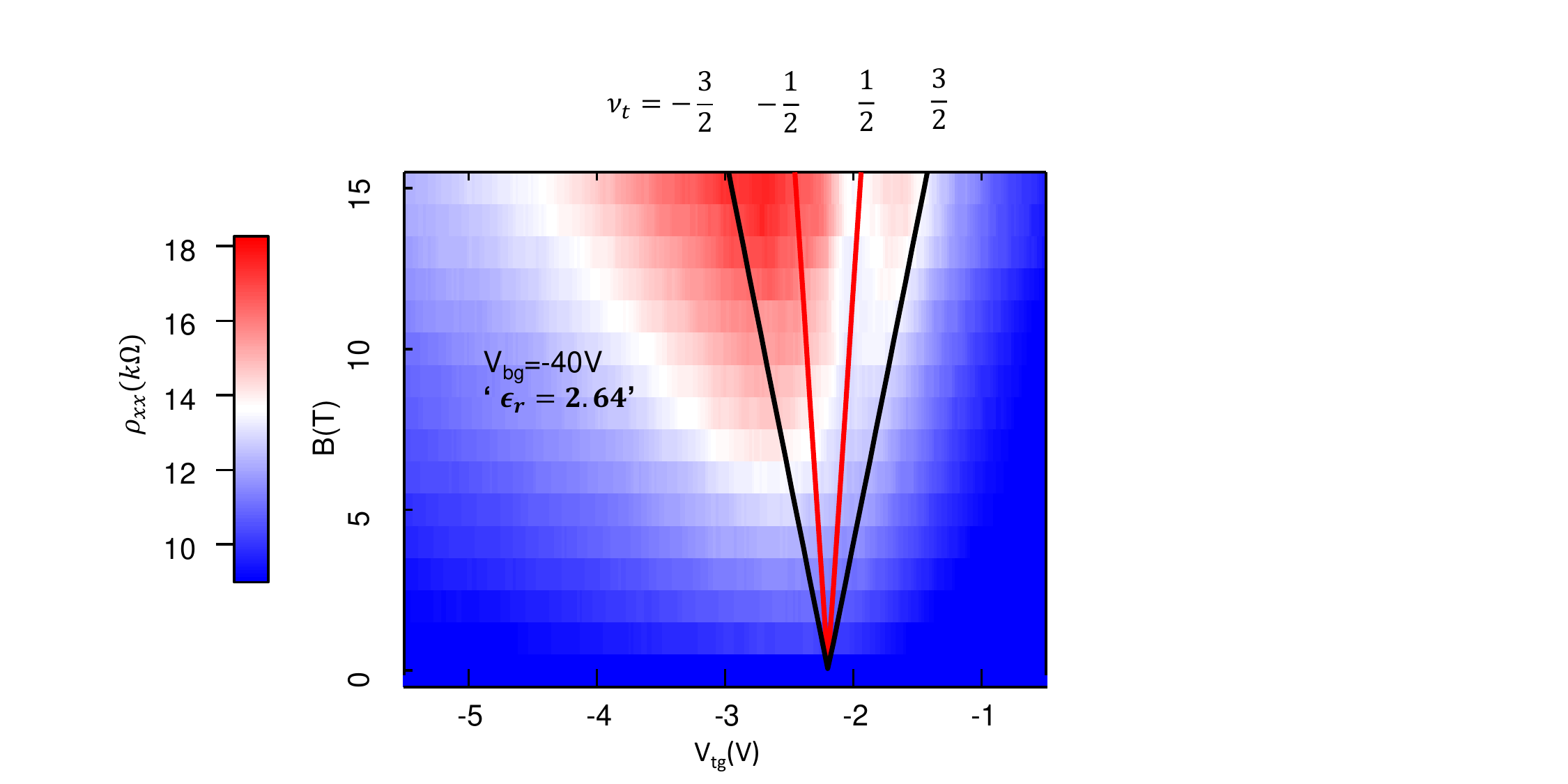}
	\caption{Linear, field-dependent LL fan diagram. The black and red lines represent the calculated LLs corresponding to a filling factor of $\pm\frac{3}{2}$ and $\pm\frac{1}{2}$}
	\label{Fig:Fig4_QHE_LLsspacing}
\end{figure}

In a Dirac cone, the energy of the LLs is given by $E_N=sgn(N)v_F\sqrt{2eB\hbar\lvert N\rvert}$. Considering that the degeneracy of the spin is lifted for the 2D topological surface states, we get a density of states of $D_{2D}(E)=\frac{\lvert E_F \rvert}{2\pi\hbar^2v_F^2}$. Using $E=\pm\hbar v_Fk$, we can deduce the number of carriers per unit area for each surface as function of the gate voltage,
\begin{equation}
N(E)=\frac{e}{h}\cdot B\cdot\lvert N\rvert=N(V_{tg})+N(V_{bg}).
\label{Equ:CarriervsVg}
\end{equation}
We found that if we use the ideal dielectric constant for the Al$_2$O$_3$ top gate and the SiO$_2$ bottom gate, the field dependence of the calculated LLs does not fit our experimental results. Therefore, we try to fit the `Landau fan diagram' using an effective dielectric constant, $\epsilon_{\textrm{eff}}$. In this way, we get a good fit with $\epsilon_{\textrm{eff}}=2.64$ for the Al$_2$O$_3$ top gate and $\epsilon_{\textrm{eff}}=1.8$ for the SiO$_2$ bottom gate, which are used in previous section of this paper to obtain boundary conditions for the two-band fits. Note that when the theoretical lines fit the fan diagram well, the spacing between different LLs should also fit the data (Fig.\ref{Fig:Fig4_QHE_LLsspacing}), confirming the chosen effective dielectric constant. The reduced dielectric constant has also been observed in previous experiments with BiSbTeSe$_2$ (see supplementary material of Ref. \onlinecite{XuQHE}).

\FloatBarrier
\subsection{7. Landauer-B\"uttiker formalism for topological insulator quantum Hall edge states}

In a Hall bar, see for example Fig. 1(a) in the main text of the paper, the edge states provide a quantized value of the Hall conductance $G_{xy}$. In a current biased Hall bar, the Hall conductance $G_{xy}$ relates to the measurable transverse resistance $R_{xy}=\frac{\mu_6-\mu_2}{I}$ and the longitudinal resistance $R_{xx}=\frac{\mu_3-\mu_2}{I}$ by $G_{xy}=\frac{R_{xy}}{R_{xx}^2+R_{xy}^2}$.

In general, to calculate device conductances from edge states, the Landauer-B\"uttiker formalism is well suited. All device terminals, assumed to be leads that are in equilibrium with potential $\mu$, are labeled with and index. The current into (positive) or out of (negative) a terminal is then given by
\begin{equation}
I_p = \sum_q G_{pq} \left(\mu_p - \mu_q \right), \label{LB}
\end{equation}
where $G_{pq}$ are the values for the edge conductance from terminal $p$ to $q$. This equation can be converted into a matrix equation that relates current, $I$, to voltage, $V=e\mu$. Putting the current values into $I$ then allows to solve for all the elements of $\mu$. With the definitions of $R_{xy}$ and $R_{xx}$ one then obtains the conductances $G_{xx}$ and $G_{xy}$. 

The top and bottom surface of a 3D topological insulator have Dirac cones with opposite helicities. Therefore, when the two surfaces are gate-tuned to both have the Fermi energy above or below the Dirac point (i.e. two electron or two hole Fermi surfaces), the edge modes of the two surfaces propagate in the same direction but are orthogonal and no scattering from one to the other is quantum mechanically allowed. Therefore, the mode conductances $\left(n_{t,b}+\frac{1}{2}\right)G_0$ add up in the elements of $G_{pq}$, e.g. $G_{12}=\left(n_t+n_b+1\right)G_0$. It is then straightforward to show that $G_{xy}=\left(n_t+n_b+1\right)G_0$. 

However, when the two surfaces of a topological insulator are gate-tuned at different sides of the Dirac point (i.e. one electron and one hole Fermi surface), the edge modes of the two surfaces are counter-propagating. In this case, the helicities of the states are equal (the sign reversal going from the top to the bottom surface is cancelled by the sign reversal going from the electron to the hole side of the Dirac cone). Here, we will derive a model for the interaction between the modes, but first we focus on the case of negligible coupling, such as is the case for a sufficiently thick topologial insulator for which the surfaces are far apart. In contrast to the QSH case which has only one mode in each direction, the counterpropagating modes in a topological insulator can consist of higher values of $n$. For example, the case of $n_t=1$ and $n_b=-1$ gives $G_{pq}$ values of $G_{12}=\left( 1+\frac{1}{2} \right)G_0$ and $G_{21}=\left(-1+\frac{1}{2} \right)G_0$, etc. Solving the Landauer-B\"uttiker equation then for a standard six-terminal Hall bar, quite surprisingly, provides a non-integer quantized value for $G_{xy}$. For example, the case with counter-propagating modes given above provides $G_{xy}=\frac{112}{73} G_0 \approx 1.53 G_0$. Quantum Hall conductance values for other filling factors are mentioned in Fig. 3(b) of the main text.

Now we take also the coupling between modes into account. We introduce a transmission parameter $\tau$ for modes that have a counterpropagating partner at the other surface. Since modes that orginate from different Landau levels are orthogonal in real space, we neglect scattering between them. As an example, we take again the case of $n_t=1$ and $n_b=-1$ case for which we then consider scattering only between the $\frac{1}{2}$ and $-\frac{1}{2}$ terms. The mode conductance at the top surface then becomes $(1+\frac{\tau}{2})G_0$ instead of $(1+\frac{1}{2})G_0$. When the electrodes are numbered from 1 to 6 clockwise around a six-terminal Hall bar, then the conductance matrix becomes
\begin{align}
G=\begin{pmatrix}
0 & 1+\frac{\tau}{2} & 0 & 0 & 0 & \frac{\tau}{2} \\ 
\frac{\tau}{2} & 0 & 1+\frac{\tau}{2} & 0 & 0 & 0 \\ 
0 & \frac{\tau}{2} & 0 & 1+\frac{\tau}{2} & 0 & 0 \\ 
0 & 0 & \frac{\tau}{2} & 0 & 1+\frac{\tau}{2} & 0 \\ 
0 & 0 & 0 & \frac{\tau}{2} & 0 & 1+\frac{\tau}{2} \\ 
1+\frac{\tau}{2} & 0 & 0 & 0 & \frac{\tau}{2} & 0
\end{pmatrix} 
G_0. \label{G}
\end{align}
Combining Eqs. (\ref{LB}) and (\ref{G}) then gives
\begin{align}
\begin{pmatrix}
I_1 \\ I_2\\I_3\\I_4\\I_5\\I_6
\end{pmatrix}
=\begin{pmatrix}
1+\tau & -1-\frac{\tau}{2} & 0 & 0 & 0 & -\frac{\tau}{2} \\ 
-\frac{\tau}{2} & 1+\tau & -1-\frac{\tau}{2} & 0 & 0 & 0 \\ 
0 & -\frac{\tau}{2} & 1+\tau & -1-\frac{\tau}{2} & 0 & 0 \\ 
0 & 0 & -\frac{\tau}{2} & 1+\tau & -1-\frac{\tau}{2} & 0 \\ 
0 & 0 & 0 & -\frac{\tau}{2} & 1+\tau & -1-\frac{\tau}{2} \\ 
-1-\frac{\tau}{2} & 0 & 0 & 0 & -\frac{\tau}{2} & 1+\tau
\end{pmatrix} G_0
\begin{pmatrix}
V_1 \\ V_2\\ V_3\\ V_4\\ V_5\\ V_6
\end{pmatrix}, \label{I}
\end{align}
where $V_p=e \mu_p$. Putting $I_1=-I_4=I$ then allows to solve the matrix equation and from which $G_{xy}$ can then be calculated as a function of $\tau$. For example, for $\tau=0.8$, as mentioned in the main text of the manuscript, a value of $G_{xy}=1.42 G_0$ is obtained. For the case of decoupled modes (in this context a thick topological insulator), there is no scattering between the modes and, therefore, $\tau=1$, giving $G_{xy}=1.53 G_0$. A very strong coupling can be modelled by taking $\tau=0$, which effectively localizes the lowest modes at the two surfaces, excluding them from the conductance. The conductance is then $G_{xy}=G_0$. Also for higher order filling factors the integer quantization is restored again for $\tau=0$, due to the cancellation of the modes. 

\subsection{8. (Anti-)symmetrization of $ \rho_{xx} $ and $\rho_{xy}$ signals in the dissipative quantum Hall regime}
In Fig. S6, we show an example for the symmetrization and anti-symmetrization as used in the main text. The $\rho_{xy}$ is not fully quantized and the $\rho_{xx}$ does not fall to zero, not even when the states are parallel-propagating. Even in the ideal case, due to the dissipative nature of the edge channels when counter-propagating, $\rho_{xx}$ can be particularly large (see section above), giving rise to a possible admixture of $ \rho_{xx} $ and $\rho_{xy}$ signals when contacts are slightly mis-aligned. Hence the need for (anti-)symmetrization in general. 

To correct the geometry effect in the Hall-bar sample, we symmetrized the $ \rho_{xx} $ signal with $\rho^{sym}_{xx}=\dfrac{\rho_{xx}(B)+\rho_{xx}(-B)}{2}$ and anti-symmetrized the $\rho_{xy}$ signal with $\rho_{xy}^{asym}=\dfrac{\rho_{xx}(B)-\rho_{xx}(-B)}{2}$. Then we use $G_{xx}=\dfrac{R_{xx}\dfrac{W}{L}}{(R_{xx}\dfrac{W}{L})^2+R_{xy}^2}$ and $G_{xy}=\dfrac{R_{xy}}{(R_{xx}\dfrac{W}{L})^2+R_{xy}^2}$ to obtain the longitudinal and transverse conductance. 

\begin{figure}
	\includegraphics[clip=true,width=17cm]{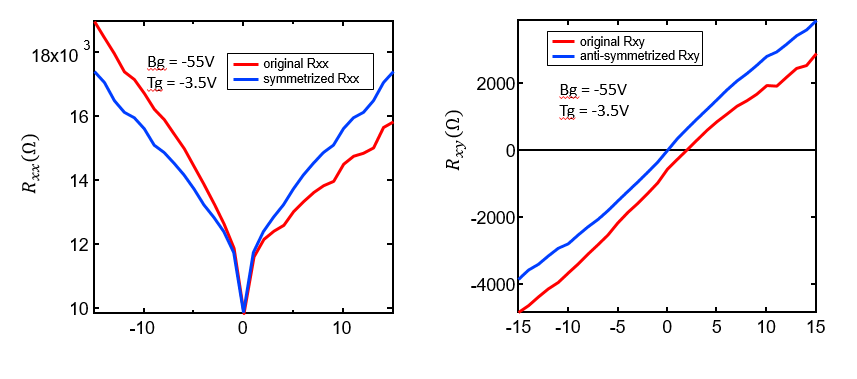}
	\caption{Symmetrization of $R_{xx}$ and anti-symmetrization of $R_{xy}$ signals. The red curves are the original data for $R_{xx}$ (left) and $R_{xy}$ (right) signal for ($V_{bg}=-55$ V,$V_{tg}=-3.5$V) . The blue curves are the symmetrized $R_{xx}$ and anti-symmetrized $R_{xy}$. }
	\label{Fig:FigS6-anti+sym}
\end{figure}